\documentclass[aps,prd,preprint,showpacs]{revtex4}
\usepackage{graphicx}
\usepackage[dvips]{pstcol}
\begin{document}
\draft
\title{The \textquotedblleft Right" Sneutrino as the LSP}
\author{A.T. ALAN$^{1}$ and S. SULTANSOY$^{2,3}$}
\address{$^{1}$Department of Physics, Abant Izzet Baysal University, 14280, Bolu, Turkey}
\address{$^{2}$ Department of Physics, Gazi University, 06500 Teknikokullar, Ankara, Turkey }
\address{$^{3}$ Institute of Physics, Academy of Sciences, H.Cavid Ave.33, Baku, Azerbaijan}
\date{\today}
\begin{abstract}
It is shown that \textquotedblleft right" sneutrino can be the
lightest supersymmetric particle. Clearly, this possibility will
drastically change decay chains of SUSY particles. The sneutrino
production at next linear colliders has been analyzed in this
scenario.
\end{abstract}
\maketitle
\section{Introduction}
R-parity conserving Minimal Supersymmetric Standard Model (MSSM)
is the most studied scenario of physics beyond the SM \cite{HG}.
Searching for supersymmetric particles is an essential part of
experimental programs of future colliders. Obviously the search
strategy is strongly dependent on mass and mixing patterns of the
SUSY particles, especially on an assumption about Lightest
Supersymmetric Particle (LSP). Usually, the lightest neutralino is
taken as the LSP. The sneutrino can be considered as an another
candidate for LSP. However, LEP1 data exclude this possibility for
\textquotedblleft left" sneutrino \cite{TH}. It should be
emphasized that this statement is valid for superpartners of
left-handed neutrinos. On the other hand, neutrino oscillation
experiments show that neutrinos have non-zero masses and it turns
out that right-handed neutrinos must exist. Thus superpartners of
the right-handed neutrinos should be included into the MSSM. As
noted in \cite{SS} the LEP1 data does not essentially constrain
the masses of superpartners of the right-handed neutrinos if their
mixings with \textquotedblleft left" sneutrinos are sufficiently
small.

In this paper we present a scenario which assumes the superpartner
of right-handed neutrino (\textquotedblleft Right" Sneutrino) to
be the Lightest Supersymmetric Particle; which is hereafter called
RS-LSP scenario.
\\In section 2 we illustrate that \textquotedblleft right"
sneutrino can be the LSP and analyze constraints on mass and
mixing of \textquotedblleft right" sneutrino coming from LEP1
data. Next in section 3 we briefly discuss decay chains of SUSY
particles in this scenario. A search for sneutrino at future
linear colliders is considered in section 4. Finally, we give some
concluding remarks in section 5.
\section{RS-LSP Scenario}
The huge number of free parameters \cite{ZSA,SD} in the three
family MSSM leads to consideration of some simplified versions,
such as the constrained MSSM ( see \cite{HE} and references
therein). In general, these simplifications ignore interfamily
mixings and possible existence of right-handed neutrinos, and
consequently their super-partners. As a result one avoids possible
conflicts with experimental data on flavor violating processes.
But at the same time we miss very interesting possible
phenomenology. In this study we deal with the three family MSSM
and do not consider possible R-parity violation, as well as GUT
and SUGRA extensions.
\subsection{MSSM with \textquotedblleft right" sneutrinos.}
There are a number of arguments favoring the existence of
right-handed neutrinos. First of all, in the framework of the SM
$\nu_\texttt{R}$'s are counterparts of the right-handed components
of the up-type quarks according to the quark-lepton symmetry.
Then, almost all extensions of the SM, with the SU(5) GUT as a
possible exception, naturally contain right-handed neutrinos.
Finally, as it was mentioned in introduction, observation of the
neutrino oscillations provides the experimental confirmation for
$\nu_\texttt{R}$. For these reasons we consider three family MSSM
with right-handed neutrinos. Therefore we deal with the six
species that constitute a family, rather than five species
considered in [5] (in the argument given below, we adopt the
notations used in this paper): $q$, $\bar{d}$, $\bar{u}$, $l$,
$\bar{e}$, and $\bar{\nu}$, where $q$ and $l$ denote weak
iso-doublets and the rest are iso-singlets. The masses of the SM
fermions and their super-partners are generated due to
\begin{eqnarray}
L=L_{scalar}+L_{Yukawa}+L_{triscalar}.
\end{eqnarray}
The first term has the form
\begin{eqnarray}
L_{scalar}= \sum_{A,i,j}m^2_{Aij}\tilde{A}_i^* \tilde{A}_j,
\end{eqnarray}
where A labels six species mentioned above, the tilde labels
sparticle and i, j=1,2,3 are family labels. Therefore this part
contains six $3\times 3$ Hermitian mass matrices. Each matrix
contains six real parameters and three phases. The Yukawa part of
the lagrangian is derived from the superpotential

\begin{eqnarray}
W_{Yukawa}=\sum_{i,j}(q_i \lambda_{uij}\bar{u}_jH_u+q_i
\lambda_{dij}\bar{d}_jH_d+l_i \lambda_{\nu ij}\bar{\nu}_jH_u+l_i
\lambda_{eij}\bar{e}_jH_d),
\end{eqnarray}
where four Yukawa matrices $\lambda$ are general $3\times 3$
matrices. Each of these matrices contains nine real parameters and
nine phases. Finally, the triscalar part is given by

\begin{eqnarray}
L_{triscalar}=\sum_{i,j}(\tilde{q}_i
a_{uij}\tilde{\bar{u}}_jH_u+\tilde{q}_i
a_{dij}\tilde{\bar{d}}_jH_d+\tilde{l}_i a_{\nu
ij}\tilde{\bar{\nu}}_jH_u+\tilde{l}_i
a_{eij}\tilde{\bar{e}}_jH_d)\times M,
\end{eqnarray}
\\where a are general $3\times 3$ matrices and M is some mass
parameter. Here in eq.~(4) in order to avoid confusion with
eq.~(3) we introduce the notations which are slightly different
from that in \cite{SD}. As a result we have 108 real parameters
(masses and mixing angles) and 90 phases. However, part of them
are unobservable because of $\texttt{U}(3)^6$ symmetry of the
gauge sector, so 18 angles and 34 phases can be rotated out and
remaining two phases correspond to baryon number in quark sector
and general lepton number. Finally, matter sector of the MSSM
(fundamental fermions and their superpartners) contains 90
observable real parameters, namely 36 masses and 54 mixing angles,
and 56 phases.
\subsection{Flavor Democracy}
Now let us recall the main assumptions of the flavor democracy
i.e., democratic mass matrix hypothesis (in the framework of the n
family SM):\\ \indent i) Before the spontaneous symmetry breaking
fermions with the same quantum numbers are indistinguishable.
Therefore Yukawa couplings are equal within each type of fermions
\cite{HHJ}: $\lambda_{uij}=\lambda_{u}$,
$\lambda_{dij}=\lambda_{d}$,
$\lambda_{lij}=\lambda_{l}$ and $\lambda_{\nu ij}=\lambda{\nu}$.\\
\indent ii) There is only one Higgs doublet, which gives Dirac
masses to all four types of fermions. Therefore Yukawa constants
for different types of fermions should be nearly equal \cite{AP}:
$\lambda_u\approx\lambda_d\approx\lambda_l\approx\lambda_\nu\approx\lambda$
. \\ \indent The first statement result in n-1 massless particles
and one massive particle with $m=n\lambda_F$ (F=u, d, l, $\nu$)
for each type of the SM fermions. The masses of the first n-1
families, as well as observable interfamily mixings, are generated
due to a small deviation from the full flavor democracy \cite{S}.
Taking into account the mass values for the third generation, the
second statement leads to the assumption that the fourth SM family
should exist. Alternatively, masses of up and down type fermions
should be generated by the interaction with different Higgs
doublets, as it takes place in the MSSM.
\subsection{\textquotedblleft Right" sneutrino as the LSP}
It is straightforward to apply flavor democracy to the MSSM. For
example according to the flavor democracy, sneutrino mass matrix
has the form
\begin{eqnarray}\left(\begin{array}{cccccc}
  m_{LL}^2 & m_{LL}^2 & m_{LL}^2 & m_{LR}^2 & m_{LR}^2 & m_{LR}^2 \\
  m_{LL}^2 & m_{LL}^2 & m_{LL}^2 & m_{LR}^2 & m_{LR}^2 & m_{LR}^2 \\
  m_{LL}^2 & m_{LL}^2 & m_{LL}^2 & m_{LR}^2 & m_{LR}^2 & m_{LR}^2 \\
  m_{RL}^2 & m_{RL}^2 & m_{RL}^2 & m_{RR}^2 & m_{RR}^2 & m_{RR}^2 \\
m_{RL}^2 & m_{RL}^2 & m_{RL}^2 & m_{RR}^2 & m_{RR}^2 & m_{RR}^2 \\
  m_{RL}^2 & m_{RL}^2 & m_{RL}^2 & m_{RR}^2 & m_{RR}^2 & m_{RR}^2 \\
\end{array}
\right)
\end{eqnarray}
\\As a result we deal with four massless
sneutrinos and two sneutrinos having the masses
\begin{eqnarray}
m^2_{3,6}=\frac{3}{2}\left(m^2_{LL}+m^2_{RR}\mp
\sqrt{(m^2_{LL}-m^2_{RR})^2+4m^2_{LR}m^2_{RL}}\right ).
\end{eqnarray}
The (small) masses of the remaining four species can be generated
due to the violation of flavor democracy. Including F- and D-term
contributions, the elements of the matrix (5) have the following
form \cite{HE}:
\begin{eqnarray}
m^2_{LL}&=&m^2_l+(\lambda_{\nu}v_u)^2+\frac{1}{2}m^2_Z \cos 2\beta,\nonumber\\
m^2_{LR}&=&m^2_{RL}=a_\nu (Mv_u-\mu v_d),\nonumber\\
m^2_{RR}&=&m^2_\nu +(\lambda_\nu v_u)^2,
\end{eqnarray}
\\where $v_u$ and $v_d$ are vacuum expectation values of the Higgs
fields $H_u$ and $H_d$, $\tan \beta =v_u/v_d$ and $\mu$ is the
supersymmetry-conserving Higgs mass parameter.\\ \indent In
\cite{TH} it is shown that LEP1 data leads to a lower bound 44.6
GeV on the sneutrino masses and, consequently in the framework of
the constrained MSSM \cite{HE}, \textquotedblleft left" sneutrino
cannot be the LSP. But LEP1 data does not essentially constrain
the masses of superpartners of the right-handed neutrinos if the
LR mixings are sufficiently small. Indeed, the contribution of the
\textquotedblleft right" sneutrino to the invisible Z width is
given by
\begin{eqnarray}
\Delta \Gamma_{inv}=0.5\times
|\delta|^2\times\left[1-\left(\frac{2\tilde{m}_\nu}{m_Z}\right)^2\right]^{3/2}\times
\Gamma_\nu
\end{eqnarray}
\\where $\delta$ denotes the \textquotedblleft left" sneutrino fraction due to
corresponding mixings and $\Gamma_\nu=167$ MeV. The experimental
value $\Delta \Gamma_{inv}\leq 2.0$ MeV leads to $|\delta|\leq
0.155$ for sufficiently light \textquotedblleft right" sneutrino.
If there are two light species one obtains
$|\delta_1|^2+|\delta_2|^2\leq 0.024$. Therefore \textquotedblleft
right" sneutrino still can be considered as the LSP both in
constrained and unconstrained MSSM.
\section{Sneutrino decays in RS-LSP scenario}
\begin{figure}
  \includegraphics[width=9cm]{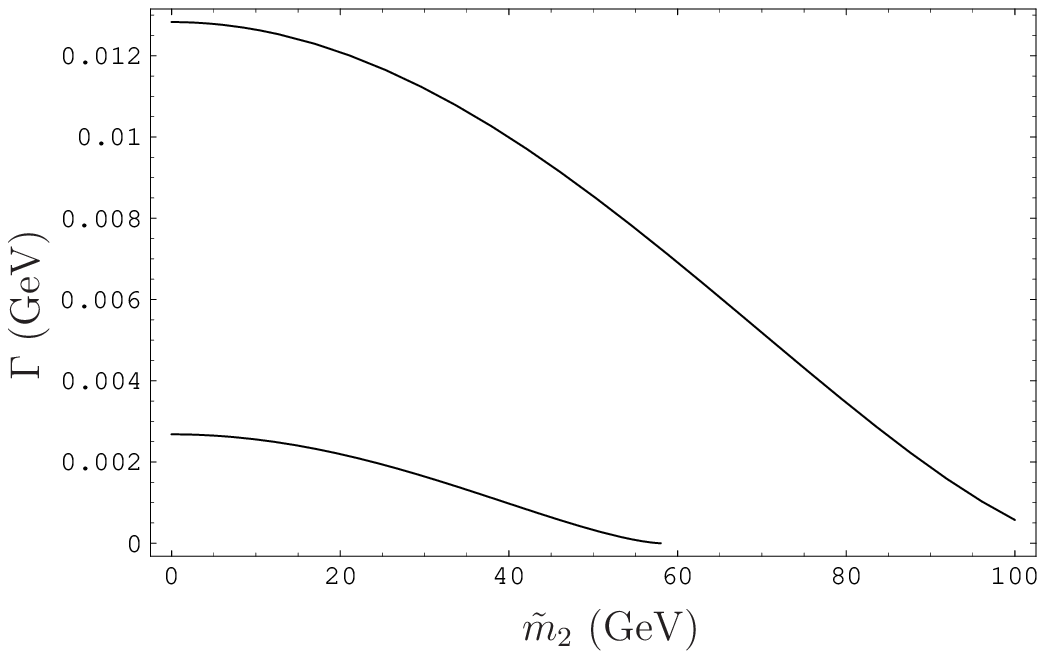}\\
  \caption{Partial decay width for $\tilde{\nu}_1^e\rightarrow
Z\tilde{\nu}_2^e$ versus $\tilde{m}_2$ with $\tilde{m}_1=200$ GeV
(upper line) and  $\tilde{m}_1=150$ GeV (lower line).}\label{1}
\end{figure}

\begin{figure}[b]
  \includegraphics[width=9cm]{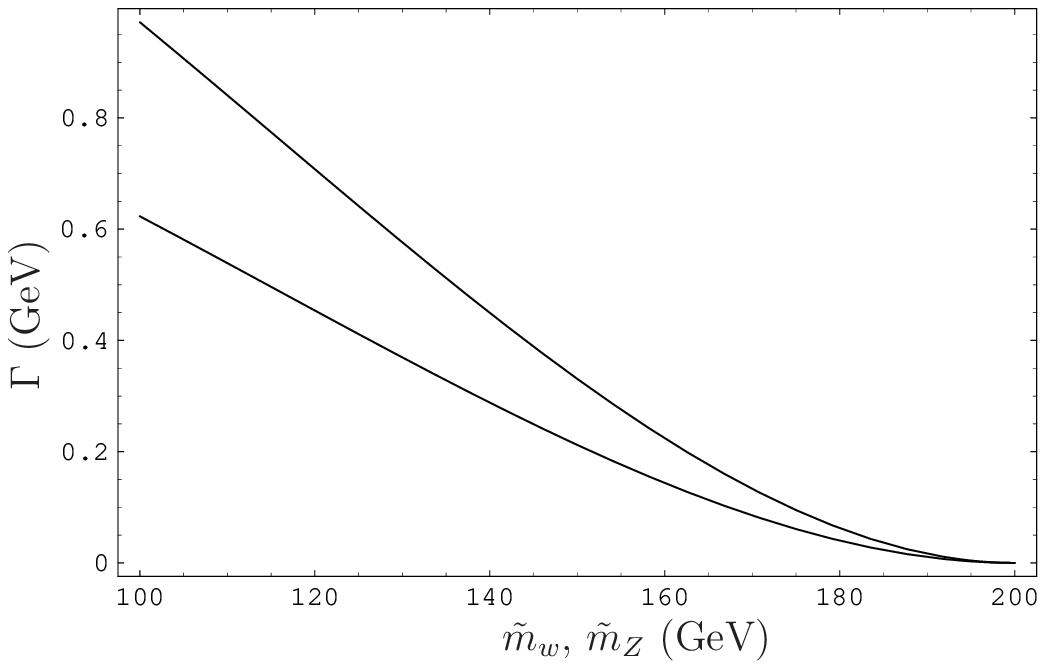}\\
  \caption{Partial decay width for $\tilde{\nu}_1^e\rightarrow
\tilde{w}e$ versus $\tilde{m}_w$ (upper line) and for
$\tilde{\nu}_1^e\rightarrow \tilde{Z}\nu_e$ versus $\tilde{m}_Z$
(lower line) with $\tilde{m}_1$=200 GeV in both cases.}\label{2}
\end{figure}

\begin{figure}
  \includegraphics[width=9cm]{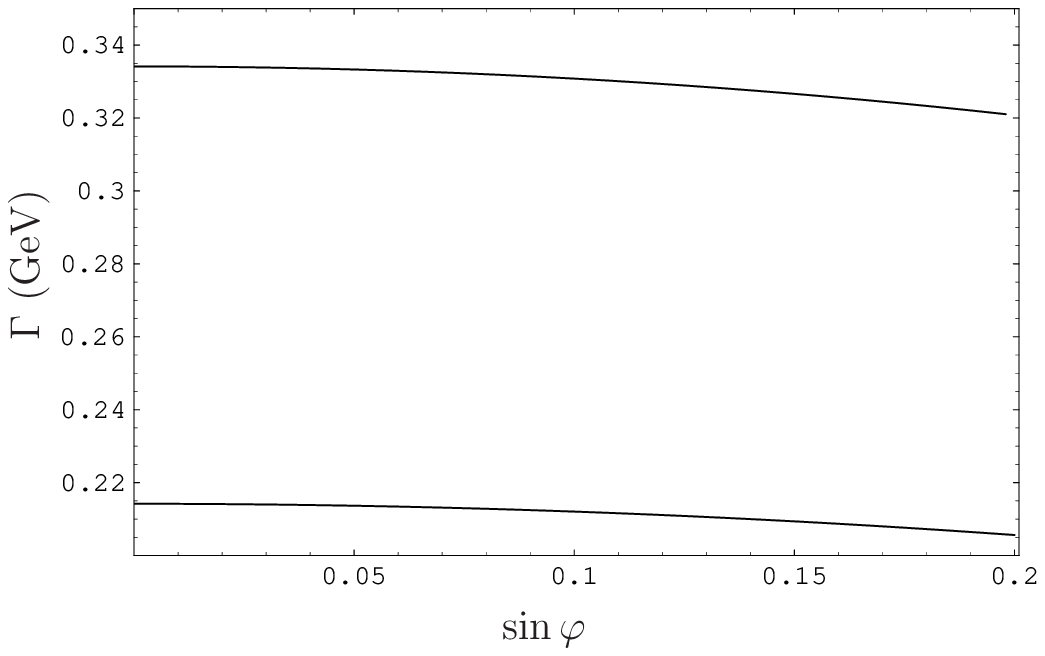}\\
  \caption{Partial decay width for $\tilde{\nu}_1^e\rightarrow
\tilde{w}e$ (upper line) and $\tilde{\nu}_1^e\rightarrow
\tilde{Z}\nu_e$ (lower line) versus $\sin\varphi$ with
$\tilde{m}_1$=200 GeV and $\tilde{m}_w$=$\tilde{m}_Z$=150 GeV.
}\label{3}
\end{figure}

Decay chains of SUSY particles depend on their masses and mixing
patterns. In general one has to deal with a $6\times6$ mass matrix
for up-type and down-type squarks, charged sleptons and
sneutrinos. Following \cite{ATS}, we ignore the interfamily
mixings so that in the sneutrino sector we are left with
\begin{eqnarray}
\tilde{\nu}_1^l=& \cos\varphi_l\tilde{\nu}_L^l + \sin \varphi_l\tilde{\nu}_R^l \nonumber\\
\tilde{\nu}_2^l=&-\sin\varphi_l\tilde{\nu}_L^l +
\cos\varphi_l\tilde{\nu}_R^l&
\end{eqnarray}
for each family. From now on $\tilde{\nu}_2^e$ is assumed to be
the LSP, so that it is stable due to R-parity conservation. In
this case the parameter $\delta$ in eq.~(8) is going to be
$\sin\varphi_e$. The mass patterns of sleptons and squarks are
assumed to satisfy
$\tilde{m}_2\ll\tilde{m}_1<\tilde{m}_{l}<\tilde{m}_{q}$, where
$\tilde{m}_2$ and $\tilde{m}_1$ denote the masses of
$\tilde{\nu}_2^e$ and $\tilde{\nu}_1^e$, respectively. If
$m_Z<\tilde{m}_1<\tilde{m}_w,\tilde{m}_Z$ the sole two body decay
mode is $\tilde{\nu}^e_1\rightarrow Z\tilde{\nu}_2^e$. For heavier
$\tilde{\nu}_1^e$, two additional decay modes
$\tilde{\nu}^e_1\rightarrow \tilde{w}e$ and
$\tilde{\nu}^e_1\rightarrow \tilde{Z}\nu_e$ do appear. In
Fig.~\ref{1} decay width $\Gamma(\tilde{\nu}_1^e\rightarrow
Z\tilde{\nu}_2^e)$ is plotted as a function of $\tilde{m}_2$ for
the following chosen values of $\tilde{m}_1=150$ GeV and 200 GeV.
Decay widths $\Gamma(\tilde{\nu}_1^e\rightarrow \tilde{w}e)$ and
$\Gamma(\tilde{\nu}_1^e\rightarrow \tilde{Z}\nu_e)$ are plotted in
Fig.~\ref{2} as functions of $\tilde{m}_w$ and $\tilde{m}_Z$ with
$\tilde{m}_1=200$ GeV. Dependences of these decay widths on
$\sin\varphi$ are shown in Fig.~\ref{3} by taking
$\tilde{m}_1=200$ GeV and $\tilde{m}_w=\tilde{m}_Z=150$ GeV. As it
is seen from Figs.~\ref{1}-\ref{3},
$\Gamma(\tilde{\nu}_1^e\rightarrow Z\tilde{\nu}_2^e)\ll
\Gamma(\tilde{\nu}_1^e\rightarrow \tilde{Z}\nu_e)<
\Gamma(\tilde{\nu}_1^e\rightarrow \tilde{w}e)$ when $\sin\varphi$
is less than 0.2 and $\tilde{m}_1-\tilde{m}_{w,Z}>10$ GeV.
\section{Sneutrino production at Future linear colliders}
\begin{figure}
  \includegraphics[width=9cm]{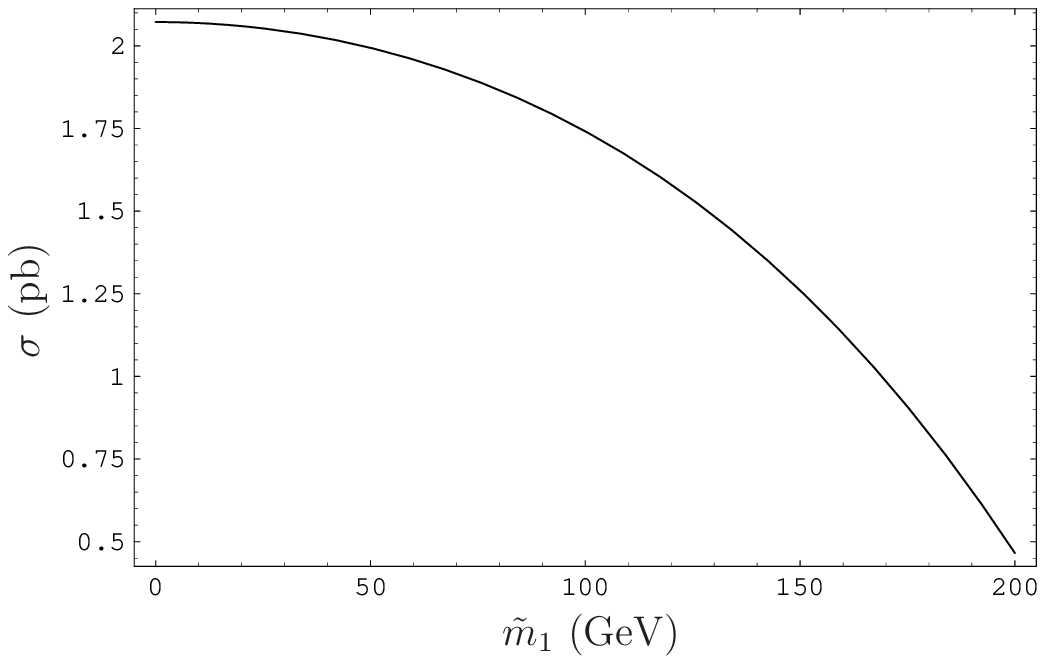}\\
  \caption{Production cross section for $e^+e^-\rightarrow
\tilde{\nu}_1^e\bar{\tilde{\nu}_1^e}$ versus $\tilde{m}_1$ at
$\sqrt{s}$=500 GeV. }\label{4}
\end{figure}

\begin{figure}
  \includegraphics[width=9.8cm]{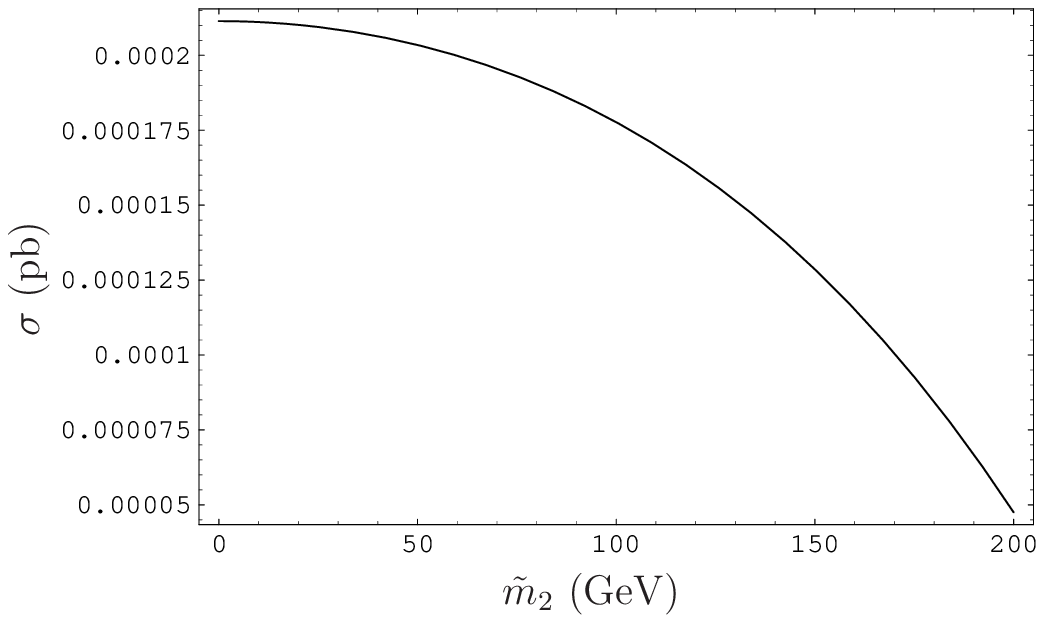}\\
  \caption{Production cross section for $e^+e^-\rightarrow
\tilde{\nu}_2^e\bar{\tilde{\nu}_2^e}$ versus $\tilde{m}_2$ at
$\sqrt{s}$=500 GeV.}\label{5}
\end{figure}

\begin{figure}
  \includegraphics[width=9cm]{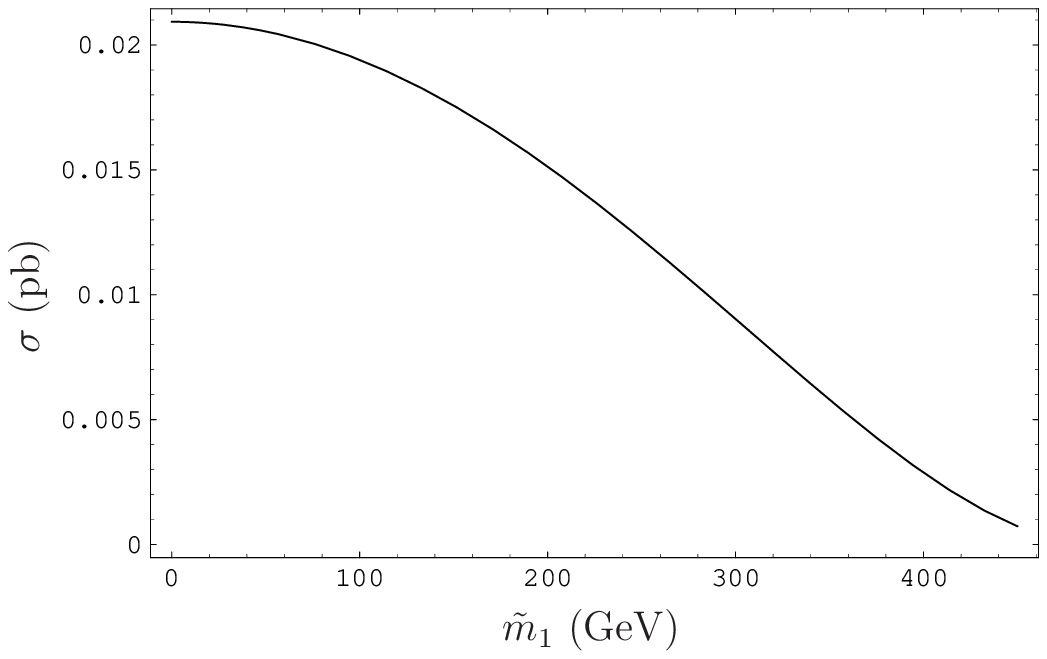}\\
  \caption{Production cross section for $e^+e^-\rightarrow
\tilde{\nu}_1^e\bar{\tilde{\nu}_2^e}$ versus $\tilde{m}_1$ with
$\tilde{m}_2=0$ at $\sqrt{s}$=500 GeV.}\label{6}
\end{figure}

\begin{figure}
  \includegraphics[width=9cm]{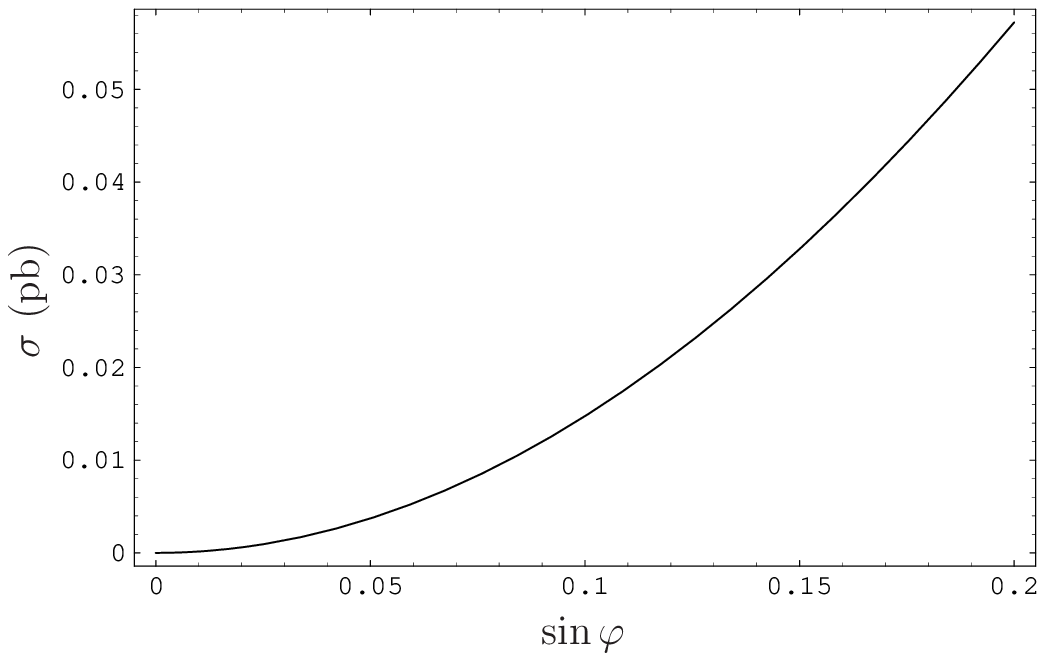}\\
  \caption{Production cross section for $e^+e^-\rightarrow
\tilde{\nu}_1^e\bar{\tilde{\nu}_2^e}$ versus $\sin\varphi$ with
$\tilde{m}_1$=200 GeV and $\tilde{m}_2$=0.}\label{7}
\end{figure}

We have four processes $e^+e^-\rightarrow
\tilde{\nu}_1^e\bar{\tilde{\nu}_1^e}$ (
$\tilde{\nu}_2\bar{\tilde{\nu}_2^e}$,
$\tilde{\nu}_1^e\bar{\tilde{\nu}_2^e}$,
$\tilde{\nu}_2^e\bar{\tilde{\nu}_1^e}$) in RS-LSP scenario instead
of one process $e^+e^-\rightarrow \tilde{\nu}_e\tilde{\nu}_e$ in
MSSM without right neutrino ( we should remember that we ignore
interfamily mixings and consider only the first lepton family).
These processes proceed via  t-channel $\tilde{w}$ exchange and
s-channel Z boson exchange. The differential cross-sections for
the processes $e^+e^-\rightarrow$
$\tilde{\nu}_1^e\bar{\tilde{\nu}_2^e}$,
$\bar{\tilde{\nu}_1^e}\tilde{\nu}_1^e$ are obtained as follows:
\begin{eqnarray}
  \frac{d\sigma(e^+e^-\rightarrow
\tilde{\nu}_1^e\bar{\tilde{\nu}_2^e},
\tilde{\nu}_2^e\bar{\tilde{\nu}_1^e})}{dt} &=&C_{12}\frac{g_W^4
[(t-\tilde{m}_1^2)(t-\tilde{m}_2^2)+ st]}{16\pi s^2}\Delta
\end{eqnarray}

where $C_{12}$=$\sin^2\varphi \cos^2\varphi$ and
\begin{eqnarray}
\Delta&=&-\frac{1}{4(t-\tilde{m}_w^2)^2}
+\frac{1-4\sin\theta_W+8\sin^2\theta_W}{16\cos^4\theta_W[(s-M_Z^2)^2+\Gamma_Z^2M_Z^2]}\nonumber\\
&-&\frac{(2\sin\theta_W-1)(s-M_Z^2)}{2\cos^2\theta_W[(s-M_Z^2)^2+\Gamma_Z^2M_Z^2](t-\tilde{m}_w^2)^2}\nonumber.
\end{eqnarray}
The differential cross-section for $e^+e^-\rightarrow$
$\tilde{\nu}_1^e\bar{\tilde{\nu}_1^e}$ is obtained from the
general expression eq.~(10) with the replacements
$C_{12}\rightarrow C_{11}=\cos^4\varphi$ and
$\tilde{m}_2\rightarrow\tilde{m}_1$ and for  $e^+e^-\rightarrow$
$\tilde{\nu}_2^e\bar{\tilde{\nu}_2^e}$ with the replacements
$C_{12}\rightarrow C_{22}=\sin^4\varphi$ and
$\tilde{m}_1\rightarrow\tilde{m}_2$. In the case of
$\sin\varphi=0$ we obtain the well-known formula for
$e^+e^-\rightarrow \tilde{\nu}_e\bar{\tilde{\nu}_e}$ \cite{HG}.
For numerical evaluations center of mass energy is taken 500 GeV,
and the values  $\tilde{m}_w=150$, $M_Z=92$ GeV,
$\sin\varphi=0.1$, $\sin^2\theta_W=0.22$, $g_W=0.66$ are used. In
Figs.~\ref{4} and \ref{5} we plot total cross sections of the
processes $e^+e^-\rightarrow \tilde{\nu}_1^e\bar{\tilde{\nu}_1^e}$
and $e^+e^-\rightarrow \tilde{\nu}_2^e\bar{\tilde{\nu}_2^e}$ as
functions of $\tilde{m}_1$ and $\tilde{m}_2$ respectively. The
cross section of the process $e^+e^-\rightarrow
\tilde{\nu}_1^e\bar{\tilde{\nu}_2^e}$ is shown in Fig.~\ref{6} as
a function of $\tilde{m}_1$ by setting $\tilde{m}_2=0$. Finally,
for illustration we give the $\sin\varphi$ dependence of
$\sigma(e^+e^-\rightarrow \tilde{\nu}_1^e\bar{\tilde{\nu}_2^e})$
in Fig.~\ref{7} by taking $\tilde{m}_1=200$ GeV and
$\tilde{m}_2=0$.

In RS-LSP scenario $\tilde{\nu}_2^e$ is stable and hence the
process $e^+e^-\rightarrow \tilde{\nu}_2^e\bar{\tilde{\nu}_2^e}$
is not observable. In principle the process $e^+e^-\rightarrow
\tilde{\nu}_2^e\bar{\tilde{\nu}_2^e}\gamma$ is observable, but it
has too small cross-section for $\sin\varphi<0.2$.

For the process $e^+e^-\rightarrow
\tilde{\nu}_1^e\bar{\tilde{\nu}_1^e}$ the most spectacular
manifestation is two Z bosons which are not coplanar, with missing
energy due to $\tilde{\nu}_1^e\rightarrow Z\tilde{\nu}_2^e$ and
$\bar{\tilde{\nu}_1^e}\rightarrow Z\bar{\tilde{\nu}_2^e}$.
Assuming these decays are dominant and considering leptonic decays
$Z\rightarrow e^+e^-$ and $Z\rightarrow \mu^+\mu^-$ we expect 700
events per year for an integrated luminosity at 100 fb$^{-1}$. For
the process $e^+e^-\rightarrow
\tilde{\nu}_1^e\bar{\tilde{\nu}_2^e}$ we consider two
possibilities. In the first $\tilde{\nu}_1^e\rightarrow
Z\tilde{\nu}_2^e$ is dominant decay mode, and in the second
$\tilde{\nu}_1^e\rightarrow \tilde{w}e$ is dominant. In the first
case we expect 90 mono-Z decaying into $e^+e^-$ and $\mu^+\mu^-$,
with large missing energy. In the second case we expect 1500
$e^+e^-$ pairs which are not coplanar, with large missing energy.
\section{Conclusion}
The RS-LSP scenario should be seriously considered as an
alternative to the neutralino-LSP scenario. Obviously in the first
case decay chains of the supersymmetric particles drastically
differ from those of the second case. Therefore the RS-LSP
scenario should be taken into account in the SUSY search programs
of future colliders. In this paper the process $e^+e^-\rightarrow
\tilde{\nu}\bar{\tilde{\nu}}$ has been analyzed. The associated
sneutrino-squark production at future lepton-hadron colliders was
considered in \cite{ATS}.
\begin{acknowledgements}
This work is partially supported by Turkish State Planning
Organization under the grant no DPT2002K120250 and Abant Izzet
Baysal University Research Found. Authors are grateful to A. Senol
and A.T. Tasci for useful discussions.
\end{acknowledgements}

\newpage



\end{document}